\documentclass[twocolumn,showpacs,amsmath,prl]{revtex4}
\usepackage{graphicx}

\begin{document}

\title{Wall ``thickness" effects on Raman spectrum shift, thermal conductivity, and
Young's modulus of single walled nanotubes}
\author{Gang Zhang}
\email{gangzh@stanford.edu} \affiliation{Department of Physics,
National University of Singapore, Singapore 117542, Republic of
Singapore} \affiliation{Department of Chemical Engineering,
Stanford University, CA 94305-4060 }
\author{Baowen Li}
\email{phylibw@nus.edu.sg} \affiliation{Department of Physics,
National University of Singapore, Singapore 117542, Republic of
Singapore} \affiliation{NUS Graduate School for Integrative
Sciences and Engineering, 117597, Republic of Singapore}
\date{4 October 2005}

\begin{abstract}
We demonstrate that at a finite temperature, an effective wall
thickness of a single walled carbon nanotube (SWNT) should be
$W=W_s+W_d$, where $W_s$ is the static thickness defined as the
extension of the outmost electronic orbit and $W_d$ the dynamic
thickness due to thermal vibration of atoms. Both molecular
simulations and a theoretical analysis show that $W_d$ is
proportional to $\sqrt{T}$. We find that the increase of dynamic
thickness with temperature is the main mechanism of Raman spectrum
shift. The introduction of the dynamic thickness changes some
conclusions about the Young's modulus and reduces the values of
thermal conductivity.
\end{abstract}

\maketitle

Nanotubes have attracted increasing attention in the last decade due to
potential applications in nanoscale electronic, mechanical and thermal
devices [1-13]. Depending on the geometrical structure, nanotubes can
exhibit fascinating properties, for example, by varying the chiral index $%
(n,m)$, nanotubes can change from semiconductors to metals. Y-K
Kwon et al.\cite{recent1} studied the thermal contraction effect
of fullerenes and carbon nanotube. They find that in nanotubes,
the gain in entropy translates into a longitudinal contraction,
which reaches a maximum at ${800K}$. Schelling et
al.\cite{recent2} studied the thermal expansion coefficient of
carbon nanotubes with empirical bond-order potential. These
studies show that the structure of carbon nanotube depends
strongly on temperature. In addition to tube radius and length,
wall thickness is another important and fundamental structure
parameter. It can influence the Young's modulus and thermal
conductivity, two typical physical quantities characterizing the
mechanical and thermal properties of nanotubes. These two
quantities depend crucially on the cross-section of the tubes.
Therefore, any ambiguity in calculating the thickness will cause
an error in these quantities, and might results in misleading
conclusions. Unfortunately, up to now, to our best knowledge, a
unified and unique way to define the thickness of nanotube does
not exist even though some discussions have been
made\cite{y-thick1, y-thick2, t-thick1, t-thick2, thick1, thick2}.
In the existing literature, a so called static thickness, defined
as the extension of the outmost electronic orbit, is used as the
wall thickness. This static thickness is temperature independent.

In this paper, we will demonstrate, with numerical evidence and
theoretical arguments, that at a finite temperature the static
thickness alone cannot be used as the wall thickness of SWNTs. One
needs to introduce a dynamic thickness to reflect the thermal
effect.

To carry out the measurement of wall thickness, a camera with
resolution on the atomic scale is supposed to be put along the
axis of SWNT. The wall thickness is determined by measuring the
width of the wall image. If the camera shutter is in femtosecond
(fs), a clear picture will be obtained, and the wall thickness
from this image is the instantaneous static thickness; but if the
shutter time is longer than fs, such as in picosecond (ps) scale,
the effect of the thermal vibrations of atoms will appear, the
wall image becomes blurry and the wall appears much thicker. We
know that in most physical process such as heat conduction along
the tube\cite{ZL04,t-thick2}, the time is much longer than ps, so
the widening of the wall thickness from thermal effect can not
simply be ignored. We call the contribution from this part
\textit{dynamic thickness} and denote it as $W_d$ . The wall
thickness should be $W=W_s+W_d$. Thus, in calculating any relevant
physical quantity, such as Young modulus and the thermal
conductivity etc, one should use the effective thickness $W$
instead of the static thickness $W_s$. We will demonstrate this
with numerical simulations that the dynamic thickness is a natural
structure parameter and affect many physical properties such as
Raman spectrum shift, Young's modulus and thermal conductivity.

We begin our discussion by studying the radial distribution
of SWNT at different temperatures. The results are obtained
from molecular dynamics (MD) simulations in which the Tersoff
empirical bond order potential\cite{Tersoff} is used. The
Hamiltonian of the carbon SWNT is:
\begin{equation}
H=\sum_i\left( \frac{p_i^2}{2m_i}+V_i\right) ,\quad V_i=\frac
12\sum_{j,j\neq i}V_{ij}
\end{equation}
where $V_{ij}=f_c(r_{ij})[V_R(r_{ij})+b_{ij}V_A(r_{ij})]$
is the Tersoff empirical bond order potential. $V_R(r_{ij})$ and $%
V_A(r_{ij})$ are the repulsive and attractive parts of the
potential, respectively, and $f_c(r)$ depending on the distance between atoms. $%
b_{ij}$ are the so-called bond parameters depending on the
bounding environment around atoms $i$ and $j$ they implicitely
contain many-body information. Tersoff potential has been used to
study thermal properties of carbon nanotubes successfully. For
detailed information, see Ref\cite{Tersoff}.

In the MD simulations, the carbon atom is treated as a mass point of zero
size. The carbon atom vibrations contain two parts: motion parallel and
perpendicular to wall. The vibrations perpendicular to the tube wall will
introduce a time dependent SWNT radius distribution. In our calculations,
three armchair-type SWNTs with different diameters, $(5,5)$, $(10,10)$ and $%
(15,15)$ are simulated, each contains 500, 1000 and 1500 atoms,
respectively. For a given temperature one million steps are used
for equilibration and over 30 000 simulation steps are used to
calculate the radial distribution. The time step is $0.8$ fs.

\begin{figure}[tbp]
\includegraphics[width=\columnwidth]{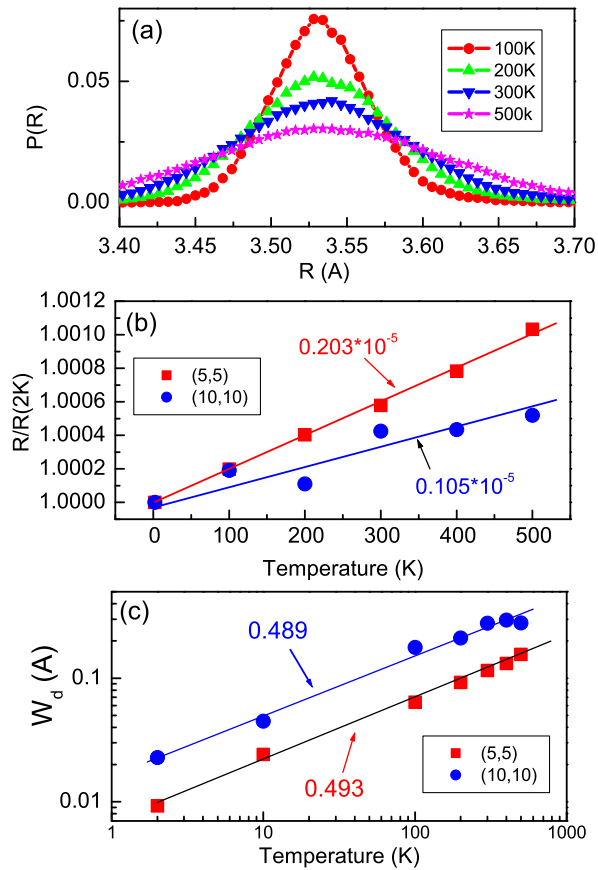} \vspace{-1cm}
\caption{(a) The radial distribution of (5,5) SWNT at different
temperatures. (b) Thermal expansion of radius for (5,5) (solid
square) and (10,10) (solid circle) SWNT. The number in the figure
is the slope of the lines obtained from best fitting. (c) The
dynamic thickness, $W_d$, versus temperature for (5,5) (solid
square) and (10,10) (solid circle). The best
fitting gives rise to: $W_d= 7\times 10^{-3}T^{0.493}$ for (5,5) tube, and $%
W_d=1.6\times 10^{-2}T^{0.489}$ for (10,10) tube.}
\label{fig:distribution}
\end{figure}

In Figure. 1, we show the radial distribution of (5,5) SWNT,
thermal expansion of the radius and the
temperature dependence of dynamic thickness for (5,5) and (10,10)
nanotubes, respectively. The radius distribution is well represented by a
Gaussian distribution. The center of the distribution is the
average radius of a carbon nanotube at the corresponding
temperature, and the width of the distribution is defined as the
dynamic thickness, $W_d$, of SWNTs. From Figure 1 we can see that
the dynamic thickness increases with temperature. The reason is
that dynamic thickness arises from the thermal vibration of atoms.
As the temperature increases, the amplitude of thermal vibration increases. $%
W_d$ versus temperature $T$ is drawn in Figure 1(c) in double
logarithmic scale, which indicates that $W_d=CT^\alpha $. For both
(5,5) and (10,10) SWNT, $\alpha \approx 0.49$, which agrees very
well with the following theoretical analysis.

In a SWNT, each carbon atom has three nearest neighbor carbon
atoms which are bond together by covalent bonds. The central
carbon atom vibrates perpendicular to the plane determined by its
three nearest neighbors (see Fig. 2). If the central atom deviates
from its equilibrium position, the force arising from its nearest
neighbors will drag it back. Within the first order approximation,
this vibration can be treated as an harmonic one with an effective
spring constant $k_e$. If the amplitude of each oscillator is
$\Delta R$, then $\langle \Delta R^2\rangle \approx 2k_BT/k_e$,
where $k_B$ is the Boltzmann constant and $T$ temperature. $k_e$
is the effective spring constant for the thermal vibration and can
be calculated as we will show below. Thus the average width caused
by the thermal vibration is
\begin{equation}
W_d\approx 2 \sqrt{\langle \Delta R^2\rangle}=2\sqrt{\frac{2k_B}{k_e}}
T^{1/2}.
\end{equation}

\begin{table}[tbp]
\caption{Dynamic thickness (second column) of three types of nanotubes at
room temperature compared with different static thickness used in different
models (columns 3-5).}
\label{tb-1}%
\begin{ruledtabular}
\begin{tabular}{ccccc}
& & &$W_d/W_s$&\\
 Tube& $W_d$ &$W_s=0.7{\AA }$\cite{y-thick1}& $W_s=1.44{\AA }$\cite{thick2}
 & $W_s=3.4{\AA }$\cite{y-thick2} \\
(5,5) &$0.115$ ${\AA }$ & $16\%$ & $8\%$ & $3\%$ \\
(10,10) & $0.278$ ${\AA }$ & $40\%$ & $20\%$ & $8\%$ \\
(15,15) & $0.304$ ${\AA }$ & $43\%$ & $21\%$ & $9\%$%
\end{tabular}
\end{ruledtabular}
\end{table}

The calculated dynamic thickness $W_d$, for three different
nanotubes at room temperature is shown in Table I. These values
are compared with the static thickness of the corresponding
nanotube used in previous studies \cite {y-thick1, y-thick2,
thick2} (columns 3-5). Dynamic thickness increases as the radius
increases, while the increase decreases for large tubes. This
arises from their different curvatures. Fig. 2 shows the detailed
structures of one carbon atom with its three nearest neighbor
atoms in (5,5) and (10,10) SWNTs. The SWNT can be seen as a
wrapped-up graphite sheet. In a graphite layer, four carbon atoms
share a common plane; while in a SWNT, the central carbon atom is
not in the same plane with its three nearest neighbors. The
distance between the atom and the plane determined by its three
nearest neighbor atoms is represented by $h$. $h$ changes with
tube diameters and is proportional to the curvature of the tube
wall, thus, the larger the tube radius the smaller $h$ becomes
(see Fig. 2). For (5,5)
SWNT, $h$ is about $0.15{\AA }$, and for (10,10) SWNT, $h$ is about $0.07{%
\AA }$. Each carbon atom in a SWNT is considered to be connected with its
nearest neighbors through the covalent bond $K^{c-c}$. $K^{c-c}$ is the effective spring
constant for the covalent bond. It is determined by the atomic interactions.
 $K^{c-c}$
is larger for strong covalent bond than for weak bond. It can be
shown approximately that the effective spring constant for the
thermal vibration is, $k_e\propto K^{c-c}h^2$. Therefore, at a
given temperature, the SWNTs with larger radius have smaller $h$,
and larger dynamic thickness.

\begin{figure}[tbp]
\includegraphics[width=\columnwidth]{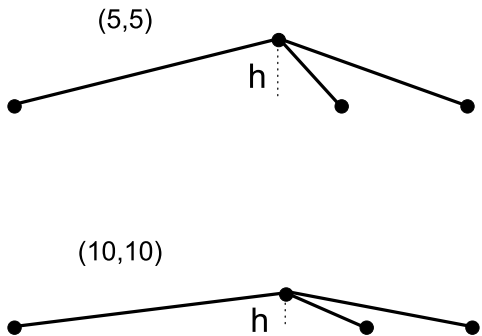} \vspace{-1cm}
\caption{Structure sketch for (5,5) and (10,10) SWNTs.}
\label{fig:structure}
\end{figure}

The $\sqrt{T}$ comes from the MD simulation results. This relation
can also be deducted from the simple harmonic model we describe
above. The perfect consistent between the simple harmonic model
and the numerical simulation demonstrate that in the temperature
range we study ($<500$K), the vibration of carbon atom in SWNT is
harmonic, this is consistent with other theoretical study that only when $T>800$K
the anharmonic effect will appear\cite{recent1}. The MD
simulations is also performed for a zigzag tube and similar effect
of dynamic thickness is found. So the dynamic thickness is a
natural structure parameter to all nanotubes. In fact, from the
analysis above, we can see that the dynamic thickness comes from
the vibrational of atom, so it exists in all types of nanotubes.

There is not a unified view on the static thickness. For example, $0.617-0.77%
{\AA }$ was used in a continuum mechanics model \cite{y-thick1, thick1}, $%
1.44{\AA }$ \cite{thick2} was used for the diameter of carbon atom, and $3.4{%
\AA }$ \cite{y-thick2}, the inter-layer separation of graphite,
was also used as the wall thickness. These different thickness are
static thickness only, while for effective thickness, one should
include the dynamic thickness. From Table I, one can see that at
room temperature, even for the smallest dynamic thickness of (5,5)
SWNT, the dynamic thickness varies between $3\%$ to $16\%$ of the
static thickness; while for bigger (15,15) SWNT, the dynamic
thickness can be as high as $10\%\sim 40\%$ of the static
thickness. Moreover, the dependence of thermal properties of SWNT
on temperature are important problems in carbon nanotubes
studies\cite{tc1, tc2, tc3, tc4}. Therefore, when a
temperature-independent static thickness is used to calculate the
thermal conductivity, the information about the temperature
dependence in thermal conductivity will be lost. The high ratio
and strong temperature dependence show that the dynamic thickness
can not be ignored in the wall thickness calculation. Ignorance of
the dynamic thickness will induce a large error in calculations of
relevant properties such as Young's modulus and thermal
conductivity. In the following we shall use the increase of the
wall thickness due to the thermal effect to explain the
\textit{Raman spectrum shift}, a correction of thermal
conductivity and Young's modulus, respectively.

\textit{Raman spectrum shift.} In high-temperature studies of Raman-active
modes in SWNTs, it is found that the Raman peak frequency shifts toward
lower frequency as temperature is increased\cite{raman1,raman2}. In an
attempt to describe this shift quantitatively, a temperature coefficient of
Raman frequency
\begin{equation}
\alpha _\omega =\frac{d\omega }{dT}\frac 1\omega
\end{equation}
was introduced\cite{raman1}. It is observed experimentally that
$\alpha _\omega =-2.47\times 10^{-5}K^{-1}$ for a SWNT with
diameter of $1.34nm$ which is close to a (10,10)
SWNT\cite{raman1}. This frequency shift has been explained as the
diameter thermal expansion of SWNTs and the softening of the
intra-tubular bonds \cite{raman1}. However, the molecular dynamics
simulation in Ref \cite{raman1} gives a much larger value of
$\alpha _\omega (=-5\times 10^{-5})$. Here we attribute the Ramman
frequency shift to the thermal expansion of the radius, $\alpha
_\omega ^R$, and the increase of the thickness, $\alpha _\omega
^{W_d}$. As we shall see, we obtain results which correspond more
closely to experiments.
\begin{figure}[tbp]
\includegraphics[width=\columnwidth]{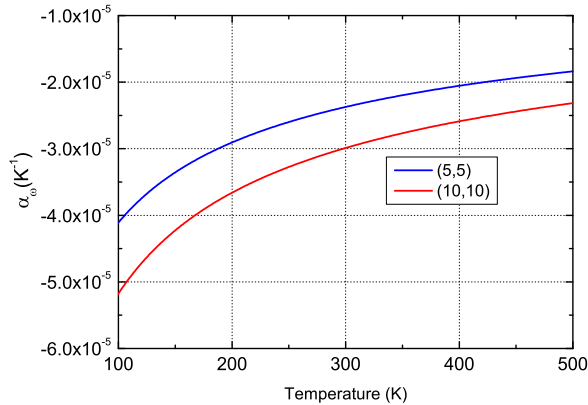} \vspace{-1cm}
\caption{Temperature coefficient of Raman frequency shift $\alpha _\omega $
versus temperature $T$ for nanotube with radius (5,5) and (10,10).}
\label{fig:AlphaOmega}
\end{figure}

After taking into account the dynamic thickness, the Raman radial breathing
frequency $\omega $ goes as $1/(R+W/2)$, where $R$ is the radius of tube and
$W$ is the wall thickness of tube. As $W_s$ is independent of temperature,
the temperature coefficient can be written as:
\begin{equation}
\alpha _\omega \approx -\frac 1{R+W/2}\left( \frac{dR}{dT}+\frac 12\frac{dW_d%
}{dT}\right) =\alpha _\omega ^R+\alpha _\omega ^{W_d}.  \label{eq:AlphaOmega}
\end{equation}
Generally speaking, the contribution from thermal expansion of the radius, $%
\alpha _\omega ^R$, is about one order of magnitude smaller than
that one from the increasing thickness $\alpha _\omega ^{W_d}$.
For example, for a (10,10) nanotube at $T=300K$, $\alpha _\omega
^R\approx -0.1\times 10^{-5}K^{-1}$ and $\alpha _\omega
^{W_d}\approx -2.84\times 10^{-5}K^{-1}$, thus we have $\alpha
_\omega \approx -2.94\times 10^{-5}K^{-1}$, which is close to the
experiment value, $-2.47\times 10^{-5}K^{-1}$. Our results show
that the temperature effect of dynamic thickness is another source
of Raman spectrum shift besides the bond softening effect.

Moreover, as $dW/dT \sim 1/\sqrt{T}$, the absolute Raman frequency shift is
larger at lower temperature and become smaller at higher temperature as is
illustrated in Fig. \ref{fig:AlphaOmega}, where the $\alpha_{\omega}$ versus
$T$ is shown. We believe that this temperature dependence behavior should be
observed in experiment.

\begin{figure}[tbp]
\includegraphics[width=\columnwidth]{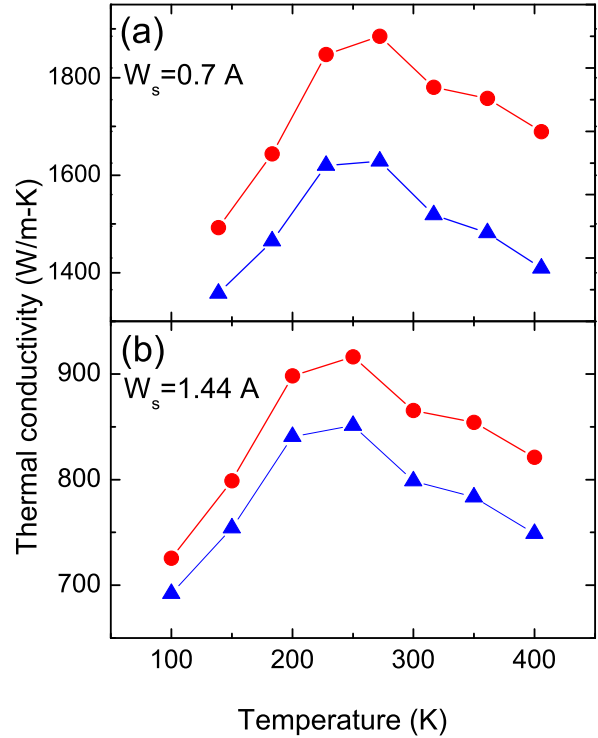} \vspace{-1cm}
\caption{Thermal conductivity for (5,5) nanotube calculated with static
thickness ($W_s$), $\kappa_e$ (solid circles), and with effective thickness (%
$W=W_s+W_d$), $\kappa$ (solid triangles). (a) Thermal conductivities
calculated with $W_s=0.7 \AA$. (b) Thermal conductivities calculated with $%
W_s=1.44 \AA$. The tube length is fixed at 50 atomic layers. The curves in
the figure are drawn to guide the eyes.}
\label{fig:kappa}
\end{figure}

\textit{Thermal conductivity}. Thermal conductivity is very sensitive to
wall thickness, as heat current is defined as energy transport through a
unit cross-section area in unit time. Different thicknesses will give rise
to different cross-sectional area of the nanotubes, thus different heat
current and thermal conductivity. The thermal conductivity, corrected after
introducing the dynamic thickness, is
\begin{equation}
\kappa=\kappa_e \left(1+\frac{W_d}{W_s}\right)^{-1}=\kappa_e \left(1+ \beta%
\sqrt{T}\right)^{-1},
\end{equation}
where $\kappa_e$ is thermal conductivity calculated with the
static wall thickness $W_s$ and $\beta=2\sqrt{2k_B/k_e}/W_s$. The
thermal conductivity, $\kappa_e$, is calculated from a
non-equilibrium molecular dynamics method, namely, two thermal
baths with slightly different temperature are put into contact
with the two ends of the nanotube, after a sufficiently time, a
stationary state is reached and a temperature gradient is set up,
the conductivity is then calculated by using, $\kappa=-JL/\Delta
T$. Where, $J$ is the heat flux, $L$ the tube length, and $\Delta
T$ the temperature difference of the two heat baths. For more
information about the calculation, please see Ref.\cite{ZL05} Thus
the real thermal conductivity is always smaller than the
experimental one. In the worst case, when $W_d/W_s\sim 0.4$, the
true thermal conductivity is approximately $70\%$ of the measured
value. In other words, the thermal conductivity in SWNTs
calculated by using static thickness, is somehow exaggerated. In
Fig. \ref{fig:kappa} we show the thermal conductivity $\kappa$ and
$\kappa_e$ for (5,5) nanotubes with different static thicknesses.
Fig \ref{fig:kappa} (a) corresponds to thermal conductivities
calculated with $W_s=0.7 \AA$ while Fig \ref{fig:kappa} (b) with
$W_s=1.44 \AA$. These calculations clearly demonstrate that the
correction of the dynamic thickness is very significant, in
particular in high temperature regime.

\textit{Young's modulus.} Another important property of nanotubes is the
Young's modulus which determines its mechanical property. In Yakobson's
calculation with (7,7) SWNT \cite{y-thick1}, $W_s=0.7{\AA }$ was used, the
Young's modulus calculated was $5.5$ Tpa; while Lu\cite{y-thick2} obtained $%
0.97$ TPa with $W_s=3.4{\AA }$. The difference can be removed if they use
the same wall thickness, thus these two Young's modulus are in fact
consistent with each other. One shall note that if we only discuss the
Young's modulus of the same tube, the wall thickness is not very important,
but if we compare the Young's modulus of different tubes, the effect of
dynamic wall thickness will be very important. For example, in Ref. \cite
{y-thick2}, by using $W_s=3.4{\AA }$ for both (5,5) and (10,10) SWNTs, with
empirical force constant model calculation, the Young's modulus for (5,5)
and (10,10) SWNT are $0.971$TPa and $0.972$TPa, respectively. From this
result, it is concluded that the Young's modulus is insensitive to the
radius of the SWNT. However, if we notice that at the same temperature, the
wall thickness is not uniform for (5,5) and (10,10) SWNT, then the
conclusion is different. By using the dynamic thickness of both tubes at
room temperature (see table I), and $W_s=3.4{\AA }$ as the static thickness,
one obtains the Young's modulus, $0.939$TPa and $0.899$ TPa for (5,5) and
(10,10), respectively. The smaller SWNT has a slightly larger Young's
modulus than the larger SWNT. If the smaller $W_s$ is used, then the dynamic
thickness correction would be even more obvious.

We would like to point out that, in experiments, it is impossible
to measure directly the thickness of SWNT. The thickness is
indirectly inferred from the measurement of other properties, such
as the Ramman frequency. The wall thickness is an ``effective
thickness", namely the dynamic thickness is already included. More
recently, a hardness transition was observed by changing pressure
on nanotube\cite{GXG}. In this study, it is obtained, from
room-temperature MD simulations, that an effective wall thickness
is about $0.66{\AA }$, which should already include finite
temperature effect. In the current paper, we split the effective
wall thickness into static thickness and dynamic thickness so as
to give the wall thickness a clear physics picture.

In summary, we have introduced a dynamic thickness, $W_d$, to
SWNTs. Unlike static thickness, which is independent of
temperature, the dynamic thickness is temperature dependent and
increases with temperature, $T$, as $\sqrt{T}$. At room
temperature, $W_d$ is comparable to the carbon atom diameter, and
thus cannot be ignored in calculating the thermal conductivity or
the Young's modulus for different tubes and/or at different
temperature. The introduction of dynamic thickness can alter
previous conclusions wherein only static thickness is used.
Moreover, we have found that the increase of dynamic thickness
with temperature is the main mechanism of the Raman spectrum
shift. Our numerical calculations agree very well with the
experiment data. The dynamic thickness has more significant
effects when the
 static thickness is small, such as in SWNTs. If the static thickness
 is large, the effects comes from dynamic thickness will decrease. And
 for certain measured properties, the effects of dynamic thickness will be
 weakened because of the time averaged in measurements.
 We believe that dynamic thickness is a natural structure
parameter with real physical effect for nano scale systems and can
be measured experimentally such as in the Ramman frequency shift.

ZG was supported by Singapore Millennium Foundation. BL was
supported partly by Faculty Research Grant of National University
of Singapore and Temasek Young Investigator Award of DSTA
Singapore under Project Agreement POD0410553.

\end{document}